\documentclass[english,preprint,showpacs]{revtex4-1}
\usepackage[latin9]{inputenc}
\usepackage{amstext, bm, hyperref, float, xcolor, color, amsmath, amsmath, graphicx, babel}
\usepackage[titletoc]{appendix}
\usepackage{hyperref}
\hypersetup{
     colorlinks = true,
     citecolor  = red,  
     linkcolor  = magenta 
}

\begin{document}
\title{Determine the chirality of Weyl fermions from the circular dichroism spectra of time-dependent ARPES}

\author{Rui Yu$^{1}$}
\email{yurui@hit.edu.cn}
\author{Hongming Weng$^{2,3}$}
\author{Zhong Fang$^{2,3}$}
\author{Hong Ding$^{2,3}$}
\author{Xi Dai$^{2,3}$}
\email{daix@iphy.ac.cn}

\affiliation{ $^{1}$Department of Physics, 
Harbin Institute of Technology, 
Harbin 150001, China}

\affiliation{$^{2}$Beijing National Laboratory for Condensed Matter Physics, 
and Institute of Physics, Chinese Academy of Sciences, 
Beijing 100190, China}

\affiliation{$^{3}$Collaborative Innovation Center of Quantum Matter, 
Beijing 100190, China}

\begin{abstract}
We show that the intensity of pumped states near Weyl point is
different when pumped with left- and right-handed circular polarized light, 
which leads to a special circular dichroism (CD) in time-dependent angle 
resolved photoemission spectra (ARPES).
We derive the expression for the CD of time-dependent ARPES, which is 
directly related to the chirality of Weyl fermions.
Based on the above derivation, we further propose a method to determine the chirality for a given Weyl point 
from the CD of time-dependent ARPES.
The corresponding CD spectra for TaAs has then been calculated from the first 
principle, which can be compared with the future experiments.

\end{abstract}

\date{\today}


\maketitle

\section{Introduction}
The research on topological quantum states has emerged as one of the major
topics in condensed matter physics in recent years.
It was ignited by the discovery of two-dimensional (2D) and 
three-dimensional (3D) topological insulators (TIs)~\cite{Hasan:2010vc,Qi:2011wt} and received significant attention
again by the discovery of 3D topological semimetals. 
In 3D topological semimetals the conduction and 
valence bands touch at certain points in the Brillouin zone and generate
nontrivial band topology~\cite{volovik_book}. Up to now, 
there are three types of topological semimetals:
Weyl semimetals, Dirac semimetals and node-line semimetals.
For Weyl semimetals, the band touching points are doubly degenerate and
distributed in the Brillouin zone as isolated points, which can be viewed
as ``magnetic monopoles'' in momentum space~\cite{ZFang2003}.
According to the so called ``no-go theorem''~\cite{no_go_theorem,Nielsen1981,Nielsen1981a}, Weyl points in lattice systems 
always appear in pairs with definite and opposite chirality.
The Dirac semimetals can be generated by overlapping two Weyl
fermions with opposite chirality at the same k-point, which has
fourfold degenerate at the band touching points and can be only 
protected by additional crystalline symmetry~\cite{Wang:2012ds,Wang:2013is}.
For node-line semimetals, the band touching points form closed loops 
in Brillouin zone around the Fermi level~\cite{NLS_Burkov_PRB_2011,NLS_weng_topological_2015,Kane_PhysRevLett_NLS,NLS_yu_topological_2015}.

The breakthrough in topological semimetals research happened 
after the material realization of Dirac semimetal states in 
Na$_3$Bi and Cd$_3$As$_2$~\cite{Wang:2012ds,Wang:2013is,PhysRevB.91.155139,PhysRevLett.113.246402,Neupane:2014kc,Liu:2014bf,Liu:2014hr}.
Starting from Dirac semimetals, one can obtain Weyl semimetals by
breaking either time-reversal~\cite{Wan:2011tp,Xu:2011dy,Burkov2011,Balents_Weyl_Physics} 
or inversion symmetry~\cite{Murakami2007,multilayerTRI,luLing_Photonics_Weyl,PhysRevB.90.155316,SeTe_PhysRevLett}.
Recently, a family of nonmagnetic and noncentrosymmetric 3D 
Weyl semimetals, the stoichiometric TaAs, TaP, NbAs and NbP, 
was first predicted theoretically~\cite{Weng:2015dy,Huang_2015ic}
and then verified experimentally~\cite{Weyl_ARPES_lv_observation_2015,Weyl_ARPES_lv_experimental_2015,yang_weyl_2015,Weyl_SdH_2015,Weyl_ARPES_xu_discovery2_2015,Weyl_ARPES_xu_discovery1_2015,DingHong_PhysRevLett2015,XuNan:ARPES}.
The intriguing expected properties characterizing the Weyl semimetal
have been checked carefully in this family of materials. 
The surface Fermi arcs have been observed in most of these materials from 
ARPES experiments~\cite{Weyl_ARPES_lv_observation_2015,Weyl_ARPES_lv_experimental_2015,Weyl_ARPES_xu_discovery1_2015,Weyl_ARPES_xu_discovery2_2015,DingHong_PhysRevLett2015,XuNan:ARPES}. 
The nontrivial $\pi$ Berry's phase has been experimentally accessed
by analyzing the Shubnikov de Haas oscillations in TaAs~\cite{Weyl_SdH_2015,zhang_sdH_2015}.
The negative magneto-resistivity due to chiral anomaly has been
observed in TaAs~\cite{Weyl_SdH_2015}. 
However all the above experiments can only prove the existence of 
the Weyl points but can not determine the chirality for each particular 
Weyl points, which is the key issue in the physics of Weyl semimetals.
In this paper, we show that the chirality of a particular Weyl point 
can be determined from the CD spectra of the time-dependent 
ARPES experiments.
%
%
%
%
%
%
%
%

\section{Weyl Hamiltonian}
The most general Hamiltonian expanded near a Weyl point can be expressed by
the following two-band model
\begin{equation}
H_{W}=\sum_{i,j}k_{i}a_{ij}\sigma_{j}=(k_{x},k_{y},k_{z}){\hat{a}}(\sigma_{x},\sigma_{y},\sigma_{z})^{T},\label{eq:H_Weyl_0}
\end{equation}
where $i,j=x,y,z$ and the matrix $a$ connects the pseudo-spin
space $\sigma$ and momentum space $\bf k$. 
Eq.~(\ref{eq:H_Weyl_0})
can be rewrite as
\begin{equation}
H_{W}=\chi\sum_{i,j}k_i\tilde{a}_{ij}\sigma_{j},\label{eq:H_Weyl_1}
\end{equation}
where $\chi=sign[Det(a)]$ is the chirality of the Weyl fermions. The
new matrix $\tilde{a}$ are defined as $\tilde{a}=\chi a$ and 
possesses positive determinant $sign[Det(\tilde{a})]=+1$. The matrix $\tilde{a}$ can be decomposed
by single value decomposition (SVD) as $\tilde{a}=S\Lambda D$, 
where $\Lambda$ is a diagonal matrix with three positive elements
denoted as $\lambda_{1}$, $\lambda_{2}$ and $\lambda_{3}$. 
$S$ and $D$ are two real orthogonal matrices 
and we can choose the gauge condition
that both of them take positive determinant, with which 
the new coordinates can be uniquely defined according to the 
``principle axises'' in the momentum and pseudo spin spaces
as $(k_{1},k_{2},k_{3})=(k_{x},k_{y},k_{z})S$ 
and $(\sigma_{1},\sigma_{2},\sigma_{3})=(\sigma_{x},\sigma_{y},\sigma_{z})D^{T}$, respectively.
%
%
%
%
%
Rotating the coordinate systems into the  principle axis defined as 
$1,2$ and $3$ in both momentum and pseudo spin spaces,
Eq.~(\ref{eq:H_Weyl_1}) can be rewritten as
\begin{equation}
H_{W}=\chi\sum_{i=1,2,3}\lambda_{i}k_{i}\sigma_{i}.\label{eq:H_Weyl_2}
\end{equation}
In Weyl semimetal materials, the Weyl points always come in pairs 
with opposite chirality. 
In the following sections, we will discuss how to detect the chirality for 
a given Weyl point from the CD of time-dependent ARPES experiment.
In the experimental setup as illustrated in Fig.~\ref{fig:setup},
the circular polarized light is shined onto a Weyl semimetal in
order to pump electrons from the occupied states to the unoccupied
states. Since the purpose is to selectively pump the 
``chiral electrons'' from the lower branch to the upper branch 
of a Weyl cone using chiral photons (see Fig.~\ref{fig:CD1})~\cite{RanYing_2015}, 
the energy of the pumping photons needs to be within the linear
range of the Weyl dispersion, which is roughly a few tens meV for TaAs, 
thus a THz source is required. The continuing-pumped unoccupied
states can then be probed by regular bulk-sensitive ARPES, 
such as soft x-ray ARPES commonly used in the study of Weyl semimetals~\cite{Weyl_ARPES_lv_observation_2015,Weyl_ARPES_lv_experimental_2015,Weyl_ARPES_xu_discovery1_2015,Weyl_ARPES_xu_discovery2_2015,DingHong_PhysRevLett2015,XuNan:ARPES}. 

\section{CD spectra of pumped states}
We start from the Hamiltonian
Eq.~(\ref{eq:H_Weyl_2}) and take $\chi=+1$ as an example to discuss
the CD spectra of pumped states for Weyl fermions.
The eigenvalues of Eq.~(\ref{eq:H_Weyl_2}) are given as
$E_{c/v}=\pm\big((\lambda_{1}k_{1})^{2}+(\lambda_{2}k_{2})^{2}+(\lambda_{3}k_{3})^{2}\big)^{1/2}$,
and the eigenfunctions for conduction and valence bands are given as
\begin{equation}
|u_c\rangle=\left[\begin{array}{c}
cos\frac{\theta_k}{2}e^{-i\frac{\phi_k}{2}}\\
sin\frac{\theta_k}{2}e^{+i\frac{\phi_k}{2}}
\end{array}\right]\;and\;|u_v\rangle=\left[\begin{array}{c}
-sin\frac{\theta_k}{2}e^{-i\frac{\phi_k}{2}}\\
+cos\frac{\theta_k}{2}e^{+i\frac{\phi_k}{2}}
\end{array}\right],\label{eq:eig_vectors}
\end{equation}
where $cos\theta_k=\frac{\lambda_{3}k_{3}}{|E|}$ and $tg\phi_k=\frac{\lambda_{2}k_{2}}{\lambda_{1}k_{1}}$.

To consider the coupling of electrons and pump light, we start from the 
microscopic Hamiltonian of an electron with spin-orbit coupling which is given by
\begin{equation}
H_{0}=\frac{\mathbf{p}^{2}}{2m}+V+\frac{\hbar}{4m^{2}}(\nabla V\times\mathbf{p})\cdot\mathbf{s},\label{eq:H_0}
\end{equation}
where \textbf{$\mathbf{p}$} is momentum operator, $V$ is crystal
potential,and $\mathbf{s}$ is electron spin operator. The Hamiltonian
for systems coupled to the electromagnetic field is obtained via the
Peierls substitution $\mathbf{p}\rightarrow\mathbf{p}-e\mathbf{A}$,
where $\mathbf{A}$ is the vector potential of the electromagnetic
filed. The electron-photon interaction term can then be obtained as
$H_{int}=H_{0}(\mathbf{p}-e\mathbf{A})-H_{0}(\mathbf{p})$ leading to
\begin{equation}
H_{int}=-e\mathbf{A}\cdot\mathbf{v},\label{eq:H_int}
\end{equation}
where 
\begin{equation}
v_{i}=\frac{\partial H_{W}}{\partial k_{i}}=\chi \lambda_{i}\sigma_{i},\label{eq:-10}
\end{equation}
is the velocity operator, where $i=1,2,3$.

\begin{figure}
\includegraphics[width=0.7\columnwidth]{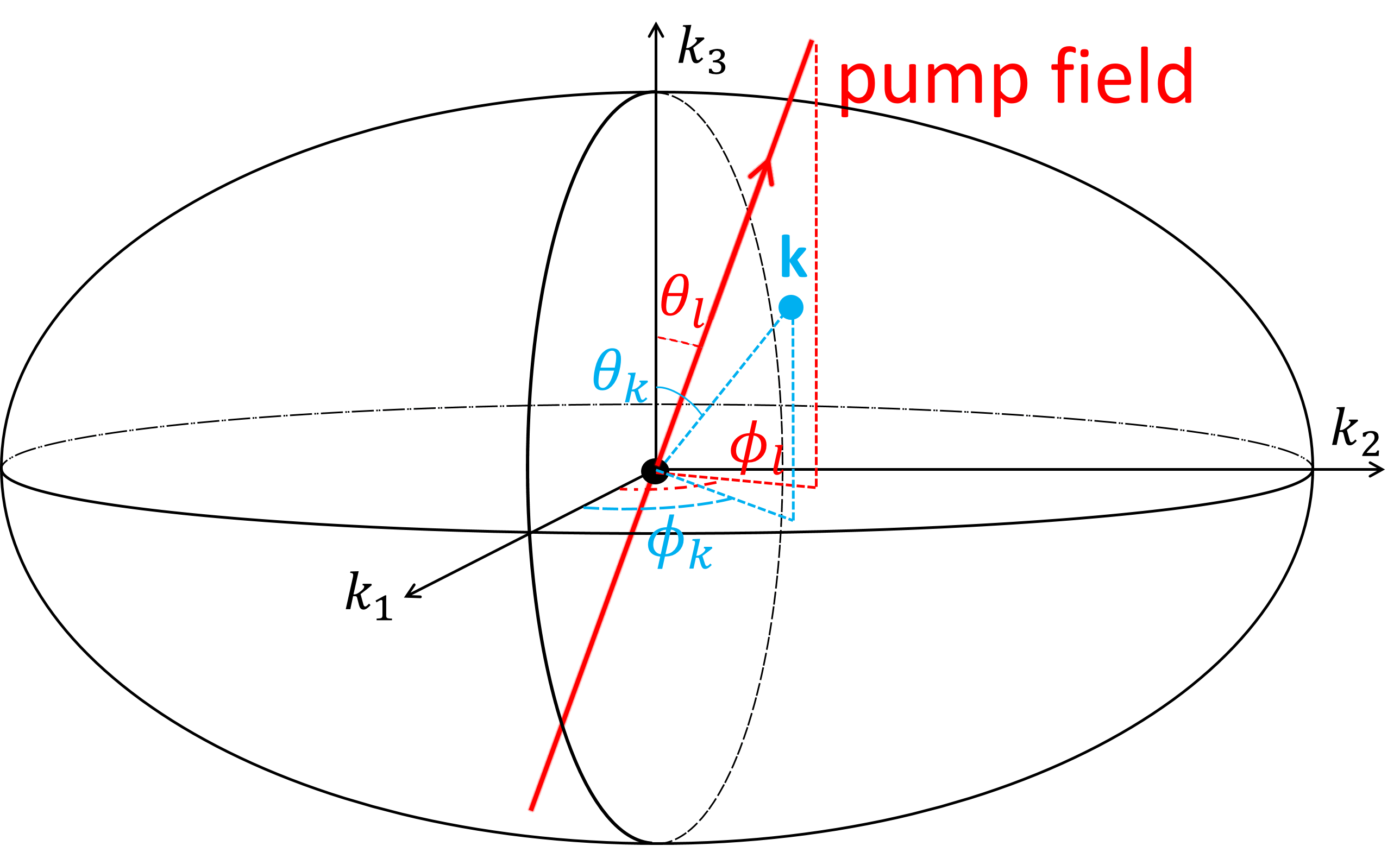}
\caption{Diagram of the experimental geometry. Circular polarized
light can be continuously rotated by $\theta _l$ and $\phi_l$ as shown in the figure.}\label{fig:setup}
\end{figure}

 Suppose in the ideal case the chemical potential is located right at the Weyl point, 
then the pumping rate from the lower branch to the upper branch
caused by the light is determined by the following matrix element
\begin{equation}
M_{cv}=\langle u_c|\mathcal{A}\cdot\mathbf{v}|u_v\rangle,\label{eq:cv_matrix}
\end{equation}
where $\mathcal{A}$ is the Fourier transform of vector potential
$\mathbf{A}$. For the circular polarized light we have
\begin{eqnarray}
\mathcal{A}_{x} & = &  A_0(cos\theta_{l}cos\phi_{l}+\eta isin\phi_{l}),\nonumber \\
\mathcal{A}_{y} & = &  A_0(cos\theta_{l}sin\phi_{l}-\eta icos\phi_{l}),\nonumber \\
\mathcal{A}_{z} & = & - A_0sin\theta_{l}.\label{eq:A}
\end{eqnarray}
where $\eta=\pm1$ indicate the right/left-handed circular polarized light 
and $\theta_{l}$ and $\phi_{l}$ are angles describing the propagating direction of the injecting light 
as illustrated in Fig.~\ref{fig:setup}.

\begin{figure}
\includegraphics[width=1\columnwidth]{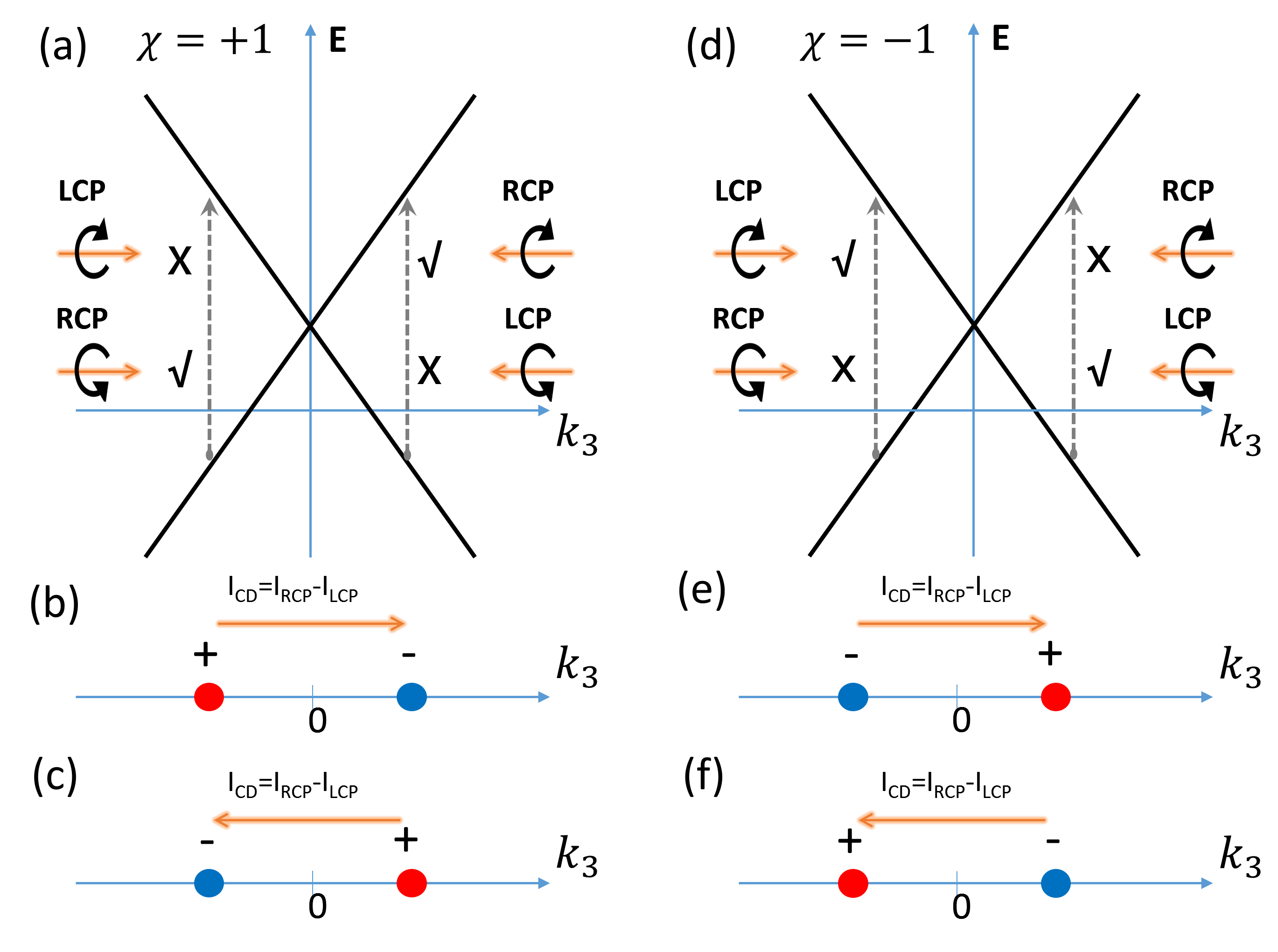}
\caption{(a) In the condition of $\lambda_{1}=\lambda_{2}$, for the Weyl fermions
with chirality $\chi=1$, the occupied states in the 
$+\hat{k}_{3}$ axis can only be pumped to the empty
states by left (right)-handed circular polarized light
injected along $\hat{k}_{3}$ ($-\hat{k}_{3}$)
direction, while for the states in the $-\hat{k}_{3}$ axis,
they can only be pumped by right (left)-circular polarized
light injected along $\hat{k}_{3}$ ($-\hat{k}_{3}$)
direction.
(b, c) The CD values with light injected along $\hat{k}_{3}$ and  $-\hat{k}_{3}$ direction for $\chi=1$.
The CD value is positive (negative) for $\bf{k}$ 
anti-parallel (parallel) to the  propagating direction 
of the light.
(d, e, f) for the Weyl fermions with chirality $\chi=-1$.
The CD value is positive (negative) for $\bf{k}$ parallel 
(anti-parallel) to the  propagating direction of the light.}
\label{fig:CD1}
\end{figure}

First we consider the simplest case with the light injecting from
one of the principal axis, say along $\hat{k}_{3}$ axis with
$\theta_{l}=0$ and $\phi_{l}=0$. 
The vector potential of light in Eq.~(\ref{eq:A}) is then simplified as 
\begin{equation}
\mathcal{A}= A_0(1,-\eta i,0),\label{eq:A_simplest}
\end{equation}
Therefore, the matrix element in Eq.~(\ref{eq:cv_matrix}) takes the formula as
\begin{equation}
M_{cv}=\langle u_c|\frac{1}{2}(\mathcal{A}_{-}v_{+}+\mathcal{A}_{+}v_{-})|u_v\rangle,\label{eq:M_cv_2}
\end{equation}
where $\mathcal{A}_{\pm}=\mathcal{A}_{1}\pm i\mathcal{A}_{2}= A_0(1\pm\eta)$
and $v_{\pm}=v_{1}\pm iv_{2}$.
For the states in the $+\hat{k}_{3}$ axis, where $\theta_k=0$, 
we get 
$|u_c\rangle_{+\hat{k}_{3}}=\left(\begin{array}{c}
1\\
0
\end{array}\right)$ and 
$|u_v\rangle_{+\hat{k}_{3}}=\left(\begin{array}{c}
0\\
1
\end{array}\right)$ as expressed in Eq.~(\ref{eq:eig_vectors}). 
To keep the discussions simple, we suppose that $\lambda_{1}=\lambda_{2}$. Then  it is easy to check that only the $v_{+}$ operator contribute nonzero matrix element between
$| u_c\rangle_{+\hat{k}_{3}}$ and $|u_v\rangle_{+\hat{k}_{3}}$ states.
Meanwhile, the requirement of $\mathcal{A_-}$ being  nonzero lead to  $\eta=-1$.
These results mean that for Weyl fermions with chirality $\chi=1$, the
occupied states in the $+\hat{k}_3$ axis can only be pumped by the 
left-handed circular polarized photons.
For the states in the $-\hat{k}_{3}$ axis, they can only be pumped by the 
right-handed circular polarized photons as illustrated in Fig.~\ref{fig:CD1}(a).
Whereas for $\chi=-1$, with the same argument as
discussed above, we find that the situation is just 
the opposite, where the states at the $\pm\hat{k}_3$
axis can only be pumped by the
right/left-handed circular polarized light as
illustrated in Fig.~\ref{fig:CD1}(d).

The pumping process induced by the circular polarized light will be 
eventually balanced by the relaxation processes in the crystal and
form a steady state.
The occupation intensity of the exited states in such a steady state
 can be obtained as~\cite{book} 
\begin{equation}
I_{\eta}
\propto|M_{cv}|^{2}A_c(E_c)A_v(E_v)
=|M_{cv}|^{2}/(\pi^2\epsilon_c\epsilon_v),\label{I_eta}
\end{equation}
where 
$A_v(\omega)=\frac{1}{\pi}\frac{\epsilon_v}{[\omega-E_v]^2+\epsilon_v^2}$ 
and
$A_c(\omega)=\frac{1}{\pi}\frac{\epsilon_c}{[\omega-E_c]^2+\epsilon_c^2}$
denote the spectral functions of the initial and final states
in the non-interaction case. The parameters $\epsilon_v$ and $\epsilon_c$ reflect the finite lifetime of electrons.
The CD spectra are defined as $I_{CD} = I_{RCP}-I_{LCP}$, 
which measures the difference 
of $I$ with right- and left-handed circular polarized pumping light.
Therefore the CD values in the $\pm\hat{k}_{3}$ axis can be calculated as
\begin{equation}
I_{CD}^{\pm\hat{k}_{3}}\propto \mp 4\chi A_0^{2}\lambda_{1}^{2}/(\pi^2\epsilon_c\epsilon_v),\label{eq:I_CD_+-kz}
\end{equation}
which shows that the CD spectra take opposite values 
at $\pm \hat{k}_{3}$ axis and its sign directly related to the chirality 
$\chi$ as shown in Fig.~\ref{fig:CD1}(b,e).

The above results indicate that we can determine the
chirality for a given Weyl point by checking the CD
of time-dependent ARPES.
But there is one problem which needs to be clarified.
In the previous paragraph, we have set the light 
propagating direction to be identical to the positive
direction of the absolution coordinate system and obtain 
the above results.
It seems that we need to know in advance that which
direction is positive (according to the gauge fixing
condition defined above) for a given principle axis,
which is not possible experimentally.
Actually this is not necessary due to the following reason.
Let's consider another possibility that the circular
polarized light is applied anti-parallel to the absolute
$\hat{k}_3$ direction.
the CD spectra can be obtained as shown in 
Fig.~\ref{fig:CD1}(c,f). 
Comparing the results in Fig.~\ref{fig:CD1}(b,c,e,f), 
we find that if we define the propagating direction 
of the circular polarized light as the reference direction,
the results for both situations become identical, 
namely for Weyl point with chirality $\chi=1$ the CD of the 
time-dependent ARPES will be positive (negative) for
momentum $\bf{k}$ anti-parallel (parallel) to the
propagating direction of light
 as illustrated in 
Fig.~\ref{fig:CD1}(b,c), while for 
$\chi=-1$ the CD spectra will be positive (negative) for
momentum $\bf{k}$ parallel (anti-parallel) to the
propagating direction of light
as illustrated in Fig.~\ref{fig:CD1}(e,f).
Therefore the above results can be used to determine the
chirality of a Weyl point purely by experiments.

For the more general case that the pump filed injected with  angles  $\theta_l$ and $\phi_l$, the CD spectra at $\bf{k}$ point with angles
$\theta_k$ and $\phi_k$ can be calculated as
\begin{eqnarray}
I_{CD} 
 & \propto & -4 \chi A_0^{2}\big(\lambda_{1}\lambda_{2}cos\theta_k cos\theta_{l}+\lambda_{3}\ sin\theta_k sin\theta_{l}\nonumber \\
 &  & *(\lambda_{2}cos\phi_k cos\phi_{l}+\lambda_{1}sin\phi_k sin\phi_{l})\big)/(\pi^2\epsilon_c\epsilon_v).\label{eq:I_CD}
\end{eqnarray}
We discuss the CD spectra for the general case in the following section.

\begin{figure}[t!]
\includegraphics[width=1\columnwidth]{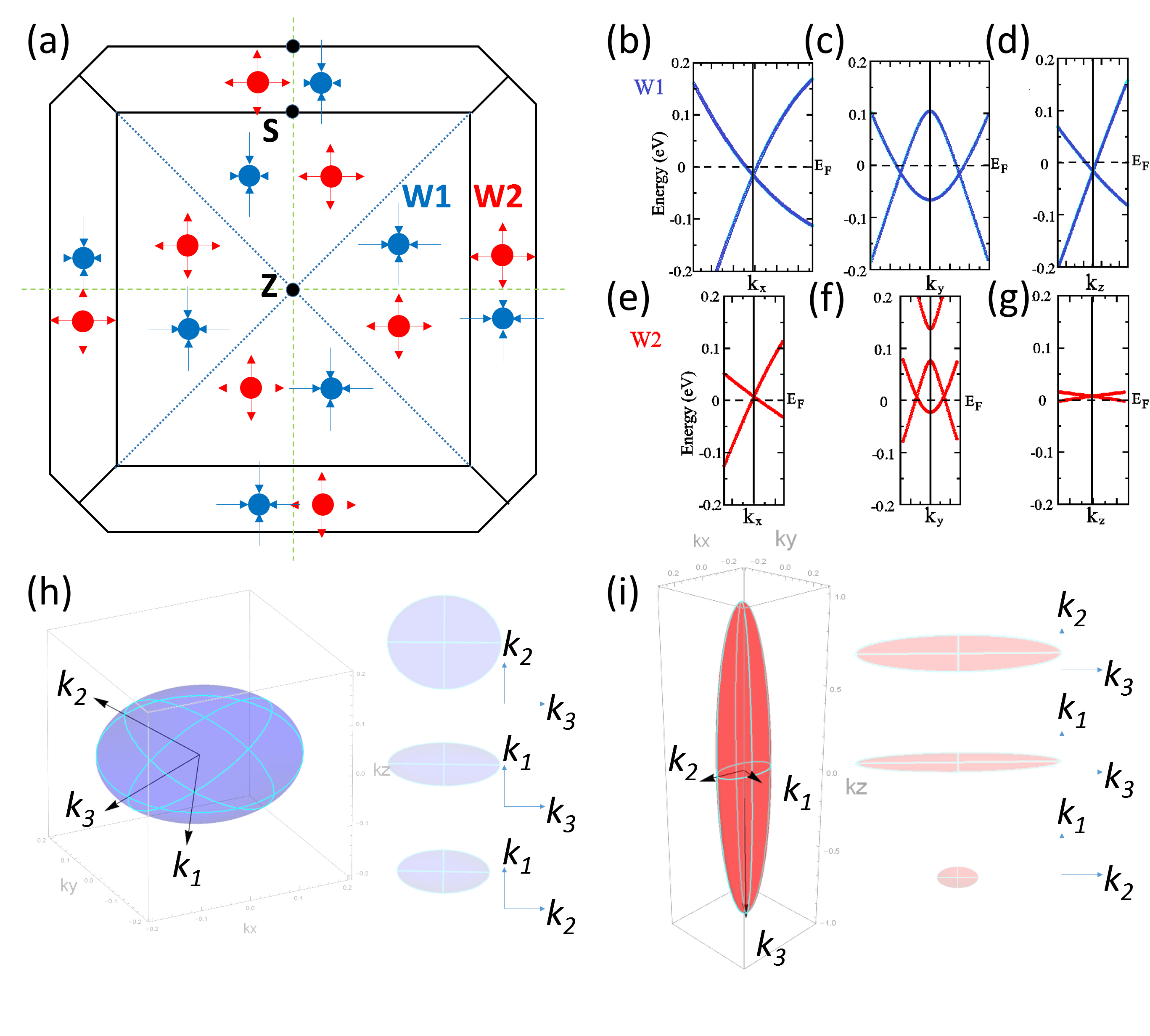}
\caption{(a) The distribution of the 12 pairs of Weyl points in BZ
for TaAs compounds. 
The red color for the Weyl point with chirality $\chi=1$ and blue 
color for $\chi=-1$. 
The two nonequivalent pairs are marked as
W1 and W2. The 4 pairs of W1 type Weyl points showing in (a) are on the top of another 4 W1 pairs, which are invisible.
The bands dispersion near W1 point and W2 point are shown in (b-d) and (e-g) respectively. (h,i) The ellipse shape isenergic surface 
near W1 and W2 points and their top view along three principle axes.
}
\label{fig:TaAs_w1w2}
\end{figure}
\section{Weyl Points in TaAs}
The new discovered TaAs family materials are the first verified realistic
materials that contain Weyl points near the Fermi level. Totally there are 12 pairs of Weyl points in TaAs 
as shown in Fig.~\ref{fig:TaAs_w1w2}(a). 
The two nonequivalent pairs are marked as 
W1 and W2. At each Weyl point, the low-energy effective Hamiltonian
takes the formula as expressed in Eq.~(\ref{eq:H_Weyl_0}).
The matrix $a$ in Eq.~(\ref{eq:H_Weyl_0}) for W1 and W2 can be
obtained by fitting with the first-principle calculations and 
expressed as
\begin{equation}
a^{W1}=\left[\begin{array}{ccc}
2.657 & -2.526 & 0.926\\
0.393 & -2.134 & 3.980\\
-1.200 & -3.530 & 1.193
\end{array}\right],\label{eq:a_W1}
\end{equation}
and 
\begin{equation}
a^{W2}=\left[\begin{array}{ccc}
1.849 & 2.531 & 0.269\\
1.849 & 1.388 & -4.910\\
0.428 & -0.302 & 0.006
\end{array}\right].\label{eq:a_W2}
\end{equation}
The energy dispersion and the isenergic surface near W1 and W2 
are shown in Fig.~\ref{fig:TaAs_w1w2}(b-i).
By performing the SVD, we can get 
the singular values of $a^{W1}$ and $a^{W2}$ as
$\lambda_1^{W1}$$=6.055$, 
$\lambda_2^{W1}$$=2.834$, 
$\lambda_3^{W1}$$=2.340$,
$\lambda_1^{W2}$$=5.564$, 
$\lambda_2^{W2}$$=2.899$ and
$\lambda_3^{W2}$$=0.520$ in unit of eV$\cdot Bohr$.
The principal axises of the 
ellipse for the isenergic surface near W1 and W2 are obtained as
$\hat{k}_{1}^{W_1}=-(0.478,0.694,0.537)$, 
$\hat{k}_{2}^{W_1}= (-0.811,0.115,0.573)$, 
$\hat{k}_{3}^{W_1}$$=(-0.336,0.710,-0.618)$,
$\hat{k}_{1}^{W_2}$$=(0.257,0.966,0.011)$, 
$\hat{k}_{2}^{W_2}$$=(0.966$$,-0.257,$$-0.007)$
 and 
$\hat{k}_{3}^{W_2}$$=(-0.004$$,0.012,$$-0.999)$ 
as shown in Fig.~{\ref{fig:TaAs_w1w2}}(h,i).

\begin{figure}
\includegraphics[width=1\columnwidth]{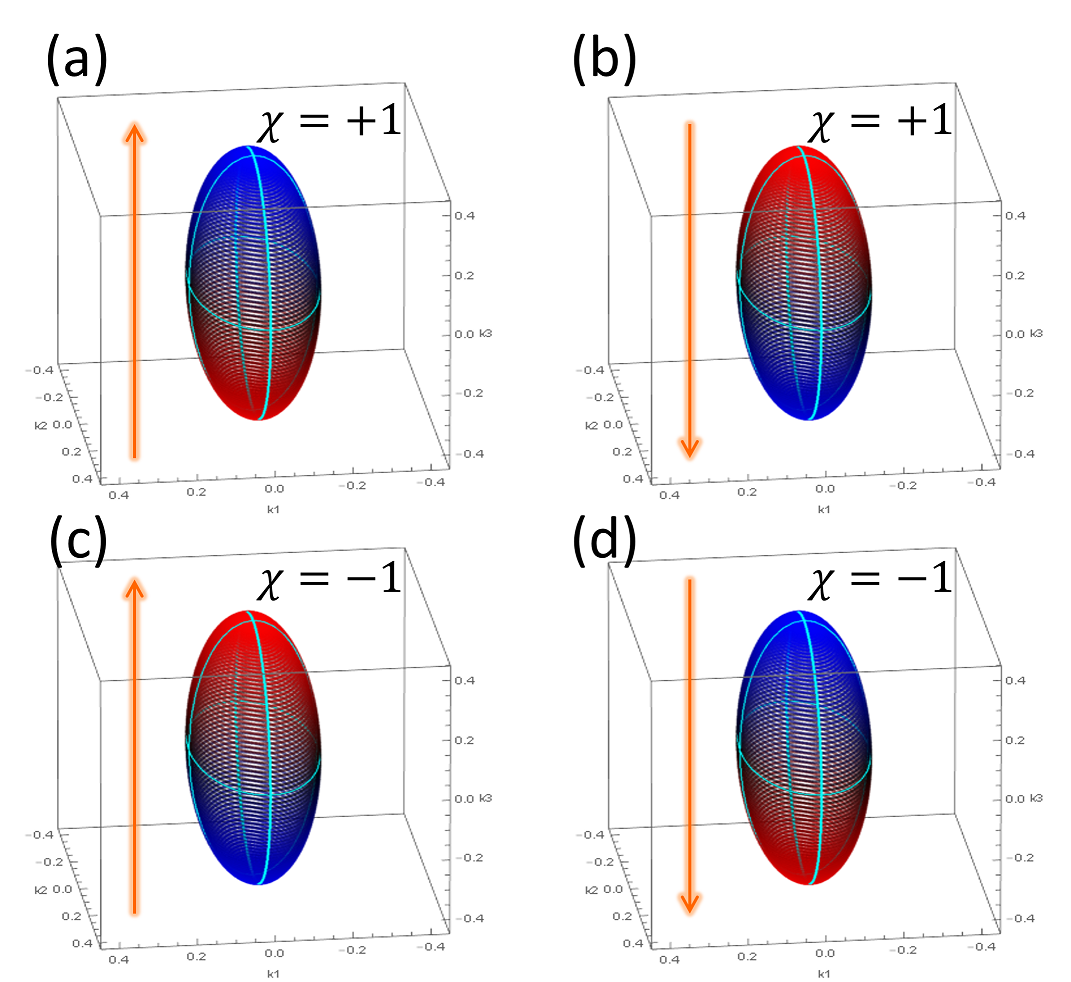}
\caption{The CD spectra for Weyl point with chirality $\chi=1$ pumped with field (a) parallel and (b) anti-parallel to $+\hat{k}_3$ principle axis.  
The red color indicate the positive CD values and blue color for the negative CD values.
In this case, the positive (negative) CD values is in the
anti-parallel (parallel) direction of the propagating direction of the light.
(c) and (d) for Weyl point with chirality $\chi=-1$. 
In this case, the positive (negative) CD values is in the 
parallel (anti-parallel) direction of the propagating direction of the light.
}
\label{fig:CD_TaAS}
\end{figure}
With the help of Eq.~(\ref{eq:I_CD}), 
The CD spectra of time-dependent ARPES for Weyl point
can be obtained as
\begin{equation}
I_{CD}\propto-4\chi A_0^{2}\lambda_{1}\lambda_{2}cos\theta_k/(\pi^2\epsilon_c\epsilon_v),\label{eq:CD_chi11}
\end{equation}
with the pump field injected along the $\hat{k}_{3}$ direction.
For the pump field injected along the $-\hat{k}_{3}$ direction, the CD spectra are  
\begin{equation}
I_{CD}\propto+4\chi A_0^{2}\lambda_{1}\lambda_{2}cos\theta_k/(\pi^2\epsilon_c\epsilon_v).\label{eq:CD_chi22}
\end{equation}

Using the criterion discussed in previous paragraph, the
positive (negative) CD values for $\bf{k}$ 
anti-parallel (parallel) to the propagating direction of 
the light indicate that $\chi=+1$,
whereas  the positive (negative) CD values for $\bf{k}$ 
parallel (anti-parallel) to the propagating direction of 
the light indicate that $\chi=-1$ as
shown in Fig.~\ref{fig:CD_TaAS}.
This criterion works even for the light propagating direction
slightly deviates from the principle axis.

%
%
%
%
%
\section{Conclusion}
In summary, we have investigated the CD of time-dependent
ARPES near Weyl points and show that the CD values is 
related to the chirality of Weyl fermions, which provide a way
to determine the chirality of the given Weyl point by
checking the sign of CD values along the propagating direction of the light.
The corresponding CD spectra for Weyl point in TaAs compound have then
been calculated, which can be compared with the future 
CD time-dependent ARPES experiments.

\noindent\textit{Acknowledgments:}
This work was supported by  the National Natural Science Foundation
of China, the 973 program of China 
(No.2013CB921700), and the ``Strategic Priority Research Program (B)" of
the Chinese Academy of Sciences (No.XDB07020100).
R.Y. acknowledges funding form the Fundamental Research Funds for 
the Central Universities (Grant No.AUGA5710059415).

\bibliographystyle{apsrev4-1}
\bibliography{refs}

\end{document}